# Embodied inquiry with AI as facilitator: an exploratory case study

**Eugenio Tufino [1]*, Paola Damiani[2]**

1 Department of Physics, Informatics and Mathematics, University of Modena and Reggio Emilia; eugenio.tufino@unimore.it
2 Department of Education and Humanities, University of Modena and Reggio Emilia; paola.damiani@unimore.it

* Correspondence: eugenio.tufino@unimore.it

**Abstract:** Generative AI is entering science education at a time when the embodied education community is asking what such systems cannot do. Rather than asking whether AI can understand the body, this exploratory case study asks where a language-based AI can stand within an inquiry activity without displacing embodied experience. University students in a Master's course in physics education investigated the statics of fluids following the ISLE approach (Investigative Science Learning Environment) in a two-phase design: students first built the buoyancy model with their own hands, without AI; a purpose-configured AI assistant then facilitated the application of the model to a new phenomenon. We discuss what a language-based facilitator cannot reach and the value of a design in which AI complements embodied inquiry rather than replacing it.

**Keywords:** embodied learning; inquiry-based learning; artificial intelligence; physics education; Universal Design for Learning

## Introduction

Generative AI is entering science education at a moment when the embodied education community is asking, with good reason, what such systems cannot do. A large language model (LLM) has no body, no senses, and no direct access to the physical world. If meaning is built through perception, action and interaction, a purely linguistic system may appear irrelevant to embodied inquiry at best, and at worst a force pulling learning back toward the verbal and the disembodied. Concerns of this kind have been raised about AI as a cognitive technology in general. Meanwhile, the systems themselves have advanced at remarkable speed: early attempts to engage chatbots in Socratic dialogue in physics showed clear limits, though current models sustain such dialogue far more competently (Gregorcic & Pendrill, 2023; Gregorcic, Polverini & Sarlah, 2024;Tufino & Gregorcic, 2025), and now answer most standard university physics assessments correctly (Kortemeyer, 2026).

Over the last decade, the complex scenario of lifelong and life wide education has posed continuous new challenges to political, school, and educational systems. This has triggered radical changes that bring both significant opportunities and serious critical issues. At the same time, research in education does not always translate into systematic changes in the daily practices of teachers and educators. Such changes should enhance the development and learning of all students, prevent neurodevelopmental disorders and socio-relational difficulties, and promote well-being and social inclusion (Damiani & Gomez Paloma, 2018). As stated by Shapiro et al. (2019), “Although embodied cognition is still in its infancy, the multidisciplinary and interdisciplinary nature of the literature provides some thought-provoking recommendations to enhance educational practice or practices, which in turn can maximize the effectiveness of the teacher in bringing about student learning” (as cited in Fiorucci, 2025, pp. 33-34). The authors saw in this literature new ways of thinking about education and classroom design, supporting the identification of the domain of Embodied Education (Damiani & Gomez Paloma, 2018; Gomez Paloma & Damiani, 2021). We have recently summarized the fundamental elements of this framework in a "Grammar of Embodied Education" to guide research and action in a coherent direction (Damiani et al., 2025).

In this study we approach the question from a different angle. Rather than asking whether AI can “understand” the body, we ask where an AI that operates through language, receiving at most the photographs of experiments students choose to send it, can stand within an inquiry activity without displacing the embodied experience that gives that inquiry its meaning. Our case comes from university physics education: a laboratory session on the statics of fluids, designed according to the Investigative Science Learning Environment (ISLE), an inquiry approach grounded in physics education research (PER) that we present in Section 2. In this session, a purpose-configured AI assistant facilitated the epistemic process of inquiry, while observation, manipulation and measurement remained entirely in the students' hands.

We present this design as an exploratory case study, at the level of feasibility rather than effectiveness. Sections 2 and 3 introduce ISLE and its reading through embodiment and multimodality; Sections 4 and 5 describe the design, the data and the exploratory findings; Section 6 discusses the boundary of a facilitator confined to language. Section 7 concludes with a transdisciplinary reflection. The article returns the two perspectives of its authors, physics education research and the pedagogy of embodiment and inclusion.

## 2. ISLE: Learning physics by doing physics

ISLE is an approach to physics teaching developed by Eugenia Etkina and collaborators over the past three decades (Etkina & Van Heuvelen, 2007; Tufino 2024). Its starting point is a simple commitment: a discipline should be learned the way it is practiced (Etkina, Brookes & Planinšič, 2019). Rather than receiving physics as a finished product to be explained and then verified, students build and test physics knowledge the way physicists do, starting from phenomena they observe with their own eyes and hands. The ISLE process is organized around three kinds of experiment (Brookes, Etkina & Planinšič, 2020). In an observational experiment, students observe a phenomenon and look for patterns, deliberately without making predictions. They then propose multiple explanations for what they saw, resisting the temptation to settle on the first idea. In a testing experiment, each explanation is put at risk: students derive a prediction in hypothetico-deductive form ("IF this explanation is correct AND we perform this experiment, THEN we should observe...") and then carry out the test. The goal is to falsify explanations, not to confirm them. Finally, in an application experiment, the surviving model is used to determine a quantity or solve a practical problem. Students work through this process in small groups, and they are expected to express their ideas in multiple representations: words, diagrams, graphs and equations (Van Heuvelen & Zou, 2001), and in practice gestures as well. Each unit, moreover, begins by creating a reason to investigate, a need to know, rather than waiting for motivation to appear (Brookes et al., 2020). For readers outside physics, the interest of ISLE lies less in its technical details than in its educational commitments. The approach is based on two intentions: that students learn to think like physicists, and that the way they learn should enhance their well-being. It rests on four foundations: the epistemology of physics, findings of cognitive science, theories of learning communities, and universal design (Brookes et al., 2020). The last two speak directly to the concerns of this special issue: knowledge is built collectively, and the environment is meant to be inclusive by design rather than by adaptation. Readers familiar with experiential learning will recognize a family resemblance with Kolb's cycle (Kolb, 1984). It is a resonance rather than a foundation: the structure of ISLE derives from how physics itself produces knowledge, not from a general theory of learning.

## 3. Where meaning lives: embodiment, multimodality, and the extended mind

Claims about embodiment in science education can mean rather different things, and it is worth declaring which meaning we rely on. Kersting et al. (2024) map three research traditions: a cognitivist one, in which bodily experience grounds conceptual structure; a phenomenological one, centered on lived experience; and an interactionist one, in which meaning is built between people as they talk, gesture and act together. For this study, our primary lens is the interactionist one, because it describes what actually happens in ISLE group work, and because it is the most demanding lens for any claim that an AI can "facilitate" inquiry. Seen through this lens, the meaning of a laboratory activity does not live inside individual heads, nor in the worksheet: it lives in the interaction. Students point, manipulate, sketch and gesture as ways of thinking together; such symbiotic gestures let them express ideas they cannot yet put into words (Gregorčič, Planinšič & Etkina, 2017; Gregorčič, 2024). Airey and Linder (2009) give this multimodal picture a sharp formulation: becoming fluent in a discipline means becoming fluent in a critical constellation of modes of disciplinary discourse, that is, words, but also diagrams, graphs, mathematics, apparatus and actions. No single mode carries the whole meaning, and some meanings are only available in some modes.

### 3.1 The inclusive perspective

Here we propose a bridge, as an interpretive inference of ours rather than an established equivalence. What physics education research calls multiple representations, and social semiotics calls multimodal resources, is what Universal Design for Learning calls multiple means (CAST, 2024). The UDL Guidelines 3.0, with their goal of learner agency and their principle of multiple means of action and expression, describe from the side of inclusion what ISLE describes from the side of physics practice: a learning environment that keeps many channels for meaning open, rather than privileging the verbal one. “Although Universal Design for Learning was originally conceived and primarily implemented in school settings, in recent years an increasingly extensive and consolidated body of scientific literature has progressively extended its application and theoretical reflection to the university context” (Fiorucci, 2025). The spread of UDL research in higher education is closely linked to the need to rethink the university as an institution open to the sociocultural changes of contemporary societies, although experience in this field is still limited (Fiorucci, 2025). Considering ISLE also presents an opportunity in this respect.

Furthermore, as discussed below, the research design focusing on ISLE entails a reflection on the limitations of AI as a facilitator: what the facilitator cannot do. This aspect is of particular interest to special and inclusive educational research, because it challenges simplistic views of digital technologies and AI as inherently inclusive tools. At the same time, it points to the need to investigate how and when these tools actually support inclusion. Where does AI belong in this scenario? The extended mind thesis holds that cognition can extend into tools and environment (Clark & Chalmers, 1998), and the direction of that extension matters. One can imagine the AI as the "brain" and humans as its "body", a direction explored provocatively for physics lab design by Samuelsson, Polverini and Gregorcic (2025). Instead, in our study the students remain the cognizing bodies, and the AI, which operates through language plus the photographs students choose to send it, enters as one more external resource they can recruit, an epistemic scaffold rather than a mind in charge. Whether such a scaffold can support ISLE inquiry without collapsing its multimodality onto the verbal channel is the empirical question of this study.

## 4. The Study: A Two-Phase Design

### 4.1. Context and design

The activity took place in a physics education research course of a Master’s degree programme at the University of Modena and Reggio Emilia, with university students, many of whom intend to become teachers, working in small groups during a laboratory session of about three hours. The session was designed so that the boundary between embodied inquiry and AI facilitation would be visible by design, in two phases. In the first phase, conducted entirely without AI, students built the physics model with their own hands. Starting from observational experiments, they explored how pressure grows with depth in water, and then, using a dynamometer to weigh objects at different depths, they constructed the model of the buoyant force: a submerged body is pushed upward by a force equal to the weight of the displaced water ($F = \varrho \cdot g \cdot V$). This is the embodied inquiry: manipulating, measuring, sketching and arguing in the group, following the ISLE sequence described in Section 2.

Only in the second phase did the AI enter, together with a new phenomenon: two sealed drink cans of identical volume, a regular cola and its sugar-free version, which behave differently when placed in water: both cans float, but the regular one floats almost completely submerged, while the sugar-free one rides visibly higher, partially emerged. Same outside, different inside: why? The students’ task was to apply and extend the model they had just built. The facilitator was a custom Gem named “Physics ISLE Lab process”, built on Google Gemini 3 Pro, configured to follow the ISLE sequence and to withhold answers. The groups interacted with it in writing, sending the photographs they chose to share, and were free to criticize its suggestions.

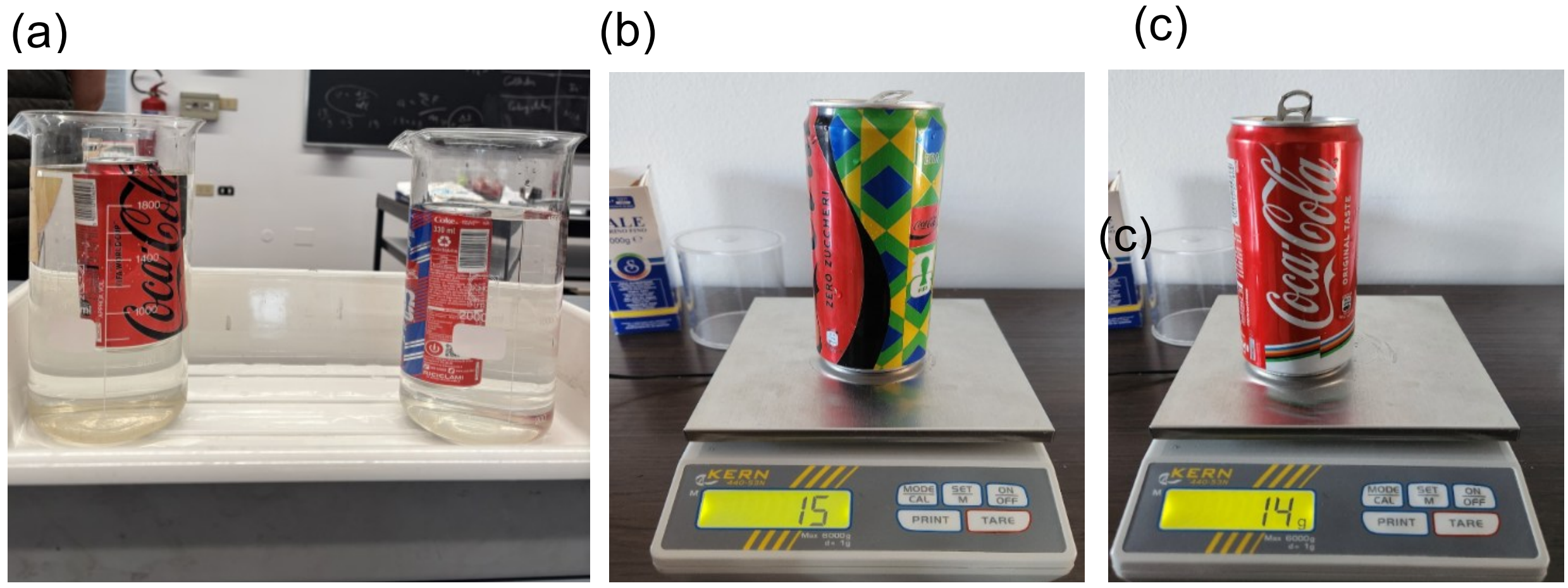

*Figure 1 (a) The two cans in water: both float, but the regular can floats almost completely submerged, while the sugar-free one rides visibly higher. (b) and (c) The two empty cans weighed on a kitchen scale: the readings, 14 g and 15 g, are compatible with the resolution of the scale (±1 g), which rules out a difference between the containers as an explanation. Photographs (b) and (c) are from a preliminary run of the activity conducted by one of the author.*

### 4.2. Data and their limits

Three pairs of students interacted with the Gem in class; a fourth participant, working alone and remotely from photographs of the phenomenon, provides a contrasting case. We anonymize them here as Pairs A, B and C, and Case D. The empirical records are the chat exports of each session. One methodological caveat applies to everything that follows: these transcripts are AI-mediated records, generated by the Gem itself as a "verbatim" rendering of the exchange, not raw interaction logs. The hard ground truth is limited to the photographs, which document the phenomenon and the identity of the cans; the numerical values that appear in the transcripts (such as the masses of the cans) are values reported by the students within AI-generated text. Our analysis is correspondingly exploratory: we read the transcripts against the ISLE sequence, asking at each prediction and measurement point whether the Gem withheld or supplied the answer.

## 5. Exploratory Findings: fragile facilitation

Across the three pairs, the Gem's first moves followed the ISLE script closely. After the groups had floated the cans and shared photographs, it named the epistemic stage and asked for at least two or three distinct explanations. The groups proposed that the sugar-free drink might be less dense, the cans' material might differ, or the amount of gas inside might differ. The Gem then asked the groups to design testing experiments and to commit to predictions before acting, in the hypothetico-deductive form introduced in Section 2: if the difference is the sugar, then the full regular can should weigh more than the sugar-free one; if the difference is the container, then the two empty cans should weigh differently (Figure 1b). Up to this point, the AI facilitation looked like ISLE by the book: it scaffolded the reasoning, not the answer.

The three trajectories then diverged. With Pair A the Gem stayed fully by the book, declaring each experiment type and requesting hypotheses and predictions before any action; the session, however, ended before the testing stage was reached. With Pair B the facilitation held under strain: when the pair measured first and formulated their prediction only afterwards, the Gem flagged the inverted order and restored it, and was careful to treat a passed test as "not falsified" rather than "proven". With Pair C it gave way: pressed by the students to "just tell us" which can weighed more, the Gem produced mass values for the two cans (384 g and 349 g) that no one had measured, effectively inventing an observation; the students then reported their own values (355 g and 341 g; all figures as reported by students within the AI-generated record).

Case D offers a revealing counterpoint. Working alone, remotely and on a different day, with photographs and balance readings provided by the instructor, this participant completed a full and physically correct ISLE cycle.

The epistemic structure was intact. What was missing was the embodied and social dimension described in Section 3. The inquiry was clean, but disembodied.

We present the resulting pattern as an interpretive hypothesis rather than a measured outcome. The Gem can hold the epistemic core of ISLE inquiry, but it is susceptible to failure under student pressure and tends to supply what should have been observed or measured.

**6. The Boundary: What a language-based facilitator cannot reach**

The Gem facilitated through language alone, supplemented only by the photographs the students chose to send it. The embodied and social dimension described in Section 3 remained invisible to it: the handling of the cans, the pointing and the symbiotic gestures, the shared sketches, the negotiation of meaning within the group as it acted (Gregorčič, Planinšič & Etkina, 2017). The activity itself remained embodied, since students manipulated, sealed, weighed and immersed the cans with their own hands; but the AI facilitator had access only to the verbal traces of that experience. This asymmetry has a predictable consequence: a language-based facilitator steers inquiry toward what can already be put into words. It cannot follow meaning as it moves between material and verbal modes, as when a group turns a manipulation into a sketch and the sketch into a claim. If left unchecked, this pull risks narrowing precisely the multiple representations and multiple means that the hands-on ISLE approach keeps open (CAST, 2024).

This limit was part of the design. Alongside the chat, students filled in a paper worksheet meant to record “what the chat cannot”: their sketch, the group's agreed statements, and a short reflection. The boundary of the facilitator was planned from the start and made visible to the students themselves. The failure of Pair C can be read in the same frame: pressed by the students, the Gem crossed the boundary between scaffolding the epistemic process and supplying the data. In an inquiry setting this cost is specific: it dissolves exactly the distinction, between what you claimed and what you measured, that the ISLE sequence is built to protect.

None of this is an argument against the design; it is an argument for its two-phase shape. The embodied inquiry came first and stayed human; the facilitator entered later, on the epistemic layer, where its channel suffices. Within this boundary, the AI is a complement to embodied inquiry, not a substitute for it.

**7. Conclusions and future work**

This exploratory case study supports three claims, all at the level of feasibility and design. First, an AI assistant configured on the ISLE approach can scaffold the epistemic core of inquiry: it named experiment types, demanded multiple explanations and predictions before tests, and in one case actively restored the epistemic order students had inverted. Second, the embodiment stays human: in a two-phase design the model is built with hands, instruments and shared meaning before the AI enters, and the language-based facilitator complements that experience rather than replacing it. Third, this facilitation is fragile, and its fidelity must be checked rather than assumed: under student pressure the Gem crossed the line from scaffolding reasoning to inventing data.

From a pedagogical point of view, the conclusions above can also be read in a transdisciplinary key: "embodiment stays human" means that the human mind-body remains central to learning and relational processes, situated within an environment that increasingly includes artificial and digital dimensions. In light of the Embodied Education perspective, AI is therefore not external to the learning environment: it is part of the relational ecosystem in which students act. What our case shows is that it is not yet part of the embodied layer itself: it enters as a mediator, on the verbal channel, while perception, manipulation and the meanings built between bodies remain human. Whether a change of mediator leaves the developmental and learning processes unchanged is an open question that this study cannot answer.

The limits are clear: one session, three pairs and one individual case, records mediated by the AI itself. What we can claim stays at the level of feasibility and design; effectiveness, and any effect on learning, can be the object of a future study. What this case already suggests, however, is a workable division of labour: the body does what only the body can do, and the AI is held, by design and by verification, to the layer where language is sufficient.

**Acknowledgements**

We thank Robin Samuelsson (Uppsala University) for valuable discussions.